\documentclass[12pt,reqno]{amsart}
\allowdisplaybreaks[4]
\usepackage{graphicx} 
\usepackage{amssymb}
\usepackage{amsmath}
\usepackage{cite}
\def\Z{{\mathbb{Z}}}

\numberwithin{equation}{section}
\makeatletter      
\@addtoreset{equation}{section}
\makeatother       

\begin{document}

\title{ Symmetric $q$-deformed KP hierarchy}

\author{Kelei Tian\dag, \  Jingsong He\ddag$^*$ \  and Yucai Su\S}
\dedicatory { \dag\ School of Mathematics,
Hefei University of technology, Hefei 230009, China \\
Email: kltian@ustc.edu.cn \\
\ddag\  Department of Mathematics, Ningbo University, Ningbo, 315211, China \\
Email: hejingsong@nbu.edu.cn \\
\S Department of Mathematics, Tongji University, Shanghai 200092, China \\
E-mail: ycsu@tongji.edu.cn
 }

\thanks{$^*$Corresponding author}

\begin{abstract} Based on  the analytic property of the symmetric $q$-exponent $e_q(x)$,  a   new  symmetric
  $q$-deformed Kadomtsev-Petviashvili ($q$-KP) hierarchy associated with the symmetric $q$-derivative operator $\partial_q$ is constructed.
Furthermore,  the symmetric $q$-CKP hierarchy and  symmetric $q$-BKP hierarchy are defined.
Here we also investigate the additional symmetries of
the  symmetric $q$-KP hierarchy.

\end{abstract}


 \maketitle



PACS(2003):\ 02.30.Ik


\section{Introduction}

The origin of $q$-calculus (quantum calculus) \cite{ks,kac}  traces
back to the early 20th century. Many mathematicians have  important
works in the area of  $q$-calculus, $q$-hypergeometric series and quantum group.
There are  two different  forms of   $q$-derivative operators,
which are  defined respectively  by

\begin{equation}
   D_q(f(x))=\frac{f(qx)-f(x)}{(q-1)x}, \qquad q\neq 1 \label{q-derivative}
   \end{equation}
and
\begin{equation}
   \partial_q(f(x))=\frac{f(qx)-f(q^{-1}x)}{(q-q^{-1})x}, \qquad q\neq 1.  \label{q-symmetrical}
   \end{equation}

The so-called $q$-deformation of the  integrable system (or $q$-deformed integrable system) started in 1990's by means of the first
$q$-derivative $D_q$  in eq.(\ref{q-derivative})   instead of usual derivative $\partial
$ with respect to $x$  in the classical system. As we know, the
$q$-deformed integrable system reduces to a classical integrable
system as $q$ goes to 1. Several $q$-deformed integrable systems have been presented, for
example, $q$-deformation of the KdV hierarchy
\cite{zhang,wuzhang,frenkel,kh}, $q$-Toda equation \cite{ZTAK},
$q$-Calogero-Moser equation \cite{ilive3}  and so on. The
$q$-deformed Kadomtsev-Petviashvili ($q$-KP) hierarchy is also a
subject of intensive study in the literature
 from \cite{mas} to \cite{linrunliang2}.
Indeed, it is worth to point out that there exist two variants of the  $q$-deformed integrable system,
one belonging to E.Frenkel \cite{frenkel} and another
 to D.H.Zhang et al. \cite{zhang,wuzhang,kh,ZTAK,ilive3,mas,iliev1,iliev2,tu,he,hetianqkp,hetianqkp2,linrunliang1,linrunliang2}.

It has been known for some time that different sub-hierarchies of the KP hierarchy can be obtained
by adding different reduction conditions on Lax operator $L$. Two
important sub-hierarchies of the KP hierarchy are CKP hierarchy \cite{dkjm2} through a
restriction $L^*=-L$ and BKP hierarchy\cite{dkjm} through a
restriction $L^*=-\partial L
\partial^{-1}$.  However, to the best of our knowledge, there is no
any results on the $q$-deformed CKP hierarchy and $q$-deformed BKP hierarchy so far.
The  difficulty  to define them
 is the conjugate operation ``$*$'' of  $q$-derivative $D_q$  in eq.(\ref{q-derivative}). In fact,
 $D_q^*\neq-D_q$  but $D_q^*=-D_q\theta^{-1}=-\frac{1}{q}D_{\frac{1}{q}}$.
This paper  shows  a quite interesting fact as $\partial_q^*=-\partial_q$,
 where the symmetric $q$-derivative operator  $\partial_q$ is defined by eq.(\ref{q-symmetrical}).
In what follows, we shall fill the gap  by constructing the   new symmetric
  $q$-deformed KP hierarchy based on the symmetric $q$-derivative operator $\partial_q$.

The paper is organized as follows. Some basic
results of symmetric $q$-derivative operator $\partial_q$ are given in Section 2, and one formula
for the symmetric $q$-exponent $e_q(x)$ is established.  Then a new symmetric $q$-KP hierarchy  are stated
 in Sections 3 similarly
 to the classical KP hierarchy \cite{dickey2}, and also
  symmetric $q$-CKP hierarchy and  symmetric $q$-BKP hierarchy are given in this section.
 We further study  the additional symmetries for the  symmetric  $q$-KP hierarchy  in Section 4.
 Section 5 is devoted to conclusions and discussions.

 \vspace{8pt}

\section{Symmetric quantum calculus}

We give   some  useful facts  about the symmetric $q$-derivative operator $\partial_q$ in the form
of eq.(\ref{q-symmetrical}) based on the literature \cite{kac}.
We work in an associative ring of functions which includes a $q$-variable $x$ and infinite time variables $t_i\in\mathbb{R}$
$$F={f=f(x; t_1, t_2, t_3,\cdots,)}.$$
The $q$-shift operator is defined by
\begin{equation}
  \theta(f(x))=f(qx).
  \end{equation}
Note that  $\theta $ does not commute with
$\partial_q$. Indeed, the relation
\begin{gather*}
(\partial_q \theta^k(f))=q^k\theta^k(\partial_q f), \qquad k\in
\mathbb{Z}
\end{gather*}
holds. The limit of $\partial_q(f(x))$ as $q$ approaches to $1$ is  the
ordinary differentiation $\partial_x(f(x)) $. We denote the formal
inverse of $\partial_q$ as $\partial_q^{-1}$.

 \vspace{8pt}
 \noindent \textbf{Proposition 1.} The conjugate of $\partial_q$ can be defined as $$\partial_q^*=-\partial_q.$$
\noindent \textbf{Proof.}  First step is to prove $\theta^*=q^{-1}\theta^{-1}$. According to the definition, we have
\begin{align*}
\partial_q(fg)&=(\theta f)(\partial_qg)+(\partial_qf)(\theta ^{-1}g) \\
&=(\theta g)(\partial_qf)+(\partial_qg)(\theta ^{-1}f).
\end{align*}
Calculating the quantum integration $\int \cdot d_qx$ for the above two formulas separately,  it follows that
\begin{align}\label{conjugateoperation1}
\int (\theta f)(\partial_qg) d_qx&=-\int(\partial_qf)(\theta ^{-1}g)  d_qx, \\ \label{conjugateoperation2}
\int (\theta g)(\partial_qf) d_qx&=-\int(\partial_qg)(\theta ^{-1}f)  d_qx.
\end{align}
Let $g\rightarrow \theta ^{-2}g$ in eq.(\ref{conjugateoperation2}), it now yields
$$\int (\theta^{-1} g)(\partial_qf) d_qx=-\int(\partial_q\theta^{-2}g)(\theta ^{-1}f)  d_qx.$$
Comparing it with the eq.(\ref{conjugateoperation1}), the above equation becomes
$$\int (\theta f)(\partial_qg) d_qx=\int(\partial_q\theta^{-2}g)(\theta ^{-1}f)  d_qx.$$
It can now be written in the form $$<\theta f, \partial_qg> = <\theta ^{-1}f, q^{-2}\theta^{-2}\partial_qg>.$$
By letting $g\rightarrow \theta ^{-2}g$ and $f\rightarrow \theta f$ in the above equation, we find that
$$<\theta^2 f, g> = <f, q^{-2}\theta^{-2}g>,$$
so one can choose $\theta^*=q^{-1}\theta^{-1}$.

We will now proceed to  prove $\partial_q^*=-\partial_q$. Let $f\rightarrow \theta^{-1} f$ and $g\rightarrow  \theta g$ in the eq.(\ref{conjugateoperation1}), it now reads
$$<\partial_q\theta^{-1}f, g> = -<f, \partial_q\theta g>.$$
This implies $$(\partial_q\theta)^*=-\partial_q\theta^{-1}.$$
According to the equation $\theta^*=q^{-1}\theta^{-1}$, we get $$\partial_q^*=-q\theta \partial_q\theta^{-1}=-\partial_q.$$
\hfill $\square$

\vspace{8pt}

The
following $q$-deformed Leibnitz rule holds
\begin{equation}
     \partial_q^n \circ f=\sum_{k\ge0}\binom{n}{k}_q\theta^{n-k}(\partial_q^kf)\theta^{-k}\partial_q^{n-k},\qquad n\in
     \Z
     \end{equation}
where the $q$-number $$(n)_q=\frac{q^n-q^{-n}}{q-q^{-1}}$$ and the $q$-binomial
is introduced as
\begin{align*}
\binom{n}{0}_q&=1,\\
\binom{n}{k}_q&=\frac{(n)_q(n-1)_q\cdots(n-k+1)_q}{(1)_q(2)_q\cdots(k)_q}, \qquad n\in
\mathbb{Z}, k\in
\mathbb{Z}_+.
\end{align*}

To illustrate the $q$-deformed Leibnitz rule, the following  examples are given.

\begin{align*}
\partial_q \circ f&=\theta(f)\partial_q+(\partial_q f)\theta^{-1},\\
\partial_q^2 \circ f&= (q+q^{-1})\theta(\partial_q
f)\theta^{-1}\partial_q +\theta^2(f)\partial_q^2+(\partial_q^2 f)\theta^{-2}, \\
 \partial_q^3\circ f&=(q^2 +q^{-2}
+1)\theta(\partial_q^2f)\theta^{-2}\partial_q + (q^2 +q^{-2} +1)\theta^2(\partial_q
f)\theta^{-1}\partial^2_q + (\partial_q^3 f)\theta^{-3}+\theta^3(f)\partial_q^3,\\
 \partial_q^{-1}\circ  f&=\theta^{-1}(f)\partial^{-1}_q- \theta^{-2}(\partial_q
f)\theta^{-1}\partial^{-2}_q + \cdots + (-1)^k \theta^{-k-1}(\partial_q^k f)\theta^{-k}\partial_q^{-k-1}+\cdots .
\end{align*}

\vspace{8pt}
Using the Taylor's formula we can get the following proposition for the symmetric $q$-exponent $e_q(x)$,
which is crucial to develop the tau  function of the symmetric $q$-KP
hierarchy and to research the interaction of $q$-solitons in the future.

\vspace{8pt}

 \noindent \textbf{Proposition 2.} The $q$-exponent $e_q(x)$ is defined as
 \begin{equation} \label{qexp}
e_q(x)=\sum_{n=0}^{\infty}\dfrac{x^n}{(n)_q!},
\end{equation}
where $$(n)_q!=(n)_q (n-1)_q (n-2)_q\cdots (1)_q,$$
then the formula
 \begin{equation} \label{qexpbyexp}
e_q(x)=\exp(\sum_{k=1}^{\infty}c_kx^k)
\end{equation}
holds, where
 \begin{equation} \label{ck}
c_k=\sum_{i=1}^{k} (-1)^{i-1}\frac{1}{i} \sum\limits_{v_1+v_2+\cdots+v_i=k \atop v_1, v_2, \cdots, v_i \in
\mathbb{Z}_+}
\frac{1}{(v_1)_q! (v_2)_q! \cdots (v_i)_q!}.
\end{equation}

\noindent \textbf{Proof.} From the definition of  $e_q(x)$ and Taylor's formula, it follows that
\begin{align*}
e_q(x)&=1+\sum_{n=1}^{\infty}\dfrac{x^n}{(n)_q!} \\
&=\exp(\ln(1+\sum_{n=1}^{\infty}\dfrac{x^n}{(n)_q!})) \\
&=\exp(\sum_{i=1}^{\infty}(-1)^{i-1}\frac{1}{i}(\sum_{n=1}^{\infty}\dfrac{x^n}{(n)_q!})^i)\\
&=\exp(\sum_{k=1}^{\infty} \sum_{i=1}^{k} (-1)^{i-1}\frac{1}{i} \sum\limits_{v_1+v_2+\cdots+v_i=k \atop v_1, v_2, \cdots, v_i \in
\mathbb{Z}_+}
\frac{ x^k}{(v_1)_q! (v_2)_q! \cdots (v_i)_q!} ) \\
&=\exp(\sum_{k=1}^{\infty}c_kx^k),
\end{align*}
where $c_{k}$ is given by eq.(\ref{ck}). \hfill $\square$

\vspace{8pt}

Several explicit forms of $q$-exponent $e_q(x)$  can be written out as follows.
\begin{align*}
c_1=&1,\\
c_2=&-\frac{(q-1)^2}{2(q^2+1)},\\
c_3=&\frac{(q-1)^2(q^4-q^3-q^2-q+1)}{3(q^2+1)(q^4+q^2+1)},\\
c_4=&-\frac{(q-1)^4(q^4-q^3-2q^2-q+1)}{4(q^2-q+1)(q^6+q^4+q^2+1)},\\
c_5=&\frac{(q-1)^4(q^{14}-2q^{13}-2q^{11}+q^{10}-2q^{9}+5q^{8}+q^{7}+5q^{6}-2q^{5}+q^{4}-2q^{3}-2q-1)}
{5(q^2+1)(q^2-q+1)(q^6+q^4+q^2+1)(q^8+q^6+q^4+q^2+1)},\\
c_6=&-\frac{(q-1)^6((q^{12}+1)(q^{2}-3q+1)+q^{2}(q^{8}+1)(q+1)-4q^{5}(q^{3}-1)(q-1)+2q^7)}
{6(q^2-q+1)(q^6+q^4+q^2+1)(q^4-q^3+q^2-q+1)(q^8-q^7+q^6+q^2-q+1)}.
\end{align*}

 \vspace{8pt}

For the case $ D_q(f(x))=\frac{f(qx)-f(x)}{(q-1)x}$  and
$\tilde{(n)}_q=\frac{q^n-1}{q-1}$, $q$-exponent function $\tilde{e}_q(x)$ is defined as $
\tilde{e}_q(x)=\sum_{n=0}^{\infty}\dfrac{x^n}{\tilde{(n)}_q!},
$ then
 \begin{equation}
\tilde{e}_q(x)=\exp(\sum_{k=1}^{\infty}\tilde{c}_kx^k),
 \end{equation}
where
 \begin{equation}
\tilde{c}_k=\frac{(1-q)^k}{k(1-q^k)}.
 \end{equation}
Recall that the $q$-exponent function $e_q(x)$ is the eigenfunction of operator $\partial_q$, i.e.
 $$\partial_q e_q(x)=e_q(x).$$
Furthermore, from $$e_q(xz)=\sum_{n=0}^{\infty}\dfrac{(xz)^n}{(n)_q!}$$
one obtains immediately that the formula
$$\partial_q^m e_q(xz)=z^me_q(xz), m=1,2,3,\cdots,$$
which is useful to define the $q$-wave function
  of the symmetric $q$-KP hierarchy in the following section.

\vspace{8pt}

\section{Symmetric $q$-deformed KP hierarchy}

Similar to the classical KP hierarchy
\cite{dkjm,dickey2}, we will define a new symmetric $q$-deformed KP hierarchy.
The Lax operator $L$ of the  symmetric $q$-KP hierarchy is given by
\begin{equation}\label{qkplaxoperator}
L=\partial_q+ u_1 +
u_2\partial_q^{-1}+u_3\partial_q^{-2}+\cdots.
\end{equation}
where $u_i=u_i(x; t_1, t_2, t_3,\cdots,),i=1, 2, 3, \cdots $. The
corresponding Lax equation of the symmetric $q$-KP hierarchy is defined by
\begin{equation}\label{qkplaxequ}
\dfrac{\partial L}{\partial t_n}=[B_n, L], \ \ n=1, 2, 3, \cdots,
\end{equation}
where the differential part $B_n=(L^n)_+=\sum\limits_{i=0}^n
b_i\partial_q^i$ and  the integral part $(L^n)_-=L^n-(L^n)_+$.

The   first few $B_n$ and flow equations in
eq.(\ref{qkplaxequ}) for dynamical variables
$\{u_1,u_{2},u_{3},\cdots\}$ can be written out as follows.

\begin{align*}
B_1 &=\partial_q+ u_1,\\
B_2 &=\partial_q^2+v_1\partial_q+ v_0,\\
B_3 &=\partial_q^3+w_2\partial_q^2+w_1\partial_q+ w_0,
\end{align*}
where $L^2=B_2+v_{-1}\partial_q^{-1}+\cdots$ and
\begin{align*}
v_1 &= \theta(u_1)+ u_1,\\
v_0 &=(\partial_qu_1)\theta^{-1}+\theta(u_2)+u_1^2+u_2,\\
v_{-1} &=(\partial_qu_2)\theta^{-1}+\theta(u_3)+u_1u_2+u_2\theta^{-1}(u_1)+u_3, \\
w_2&=\theta(v_1)+ u_1,\\
w_1&=(\partial_qv_1)\theta^{-1}+\theta(v_0)+u_1v_1+u_2,\\
w_0&=(\partial_qv_0)\theta^{-1}+\theta(v_{-1})+u_1v_0+u_2\theta^{-1}(v_1)+u_3.
\end{align*}

\vspace{8pt}

The first flow equations are

\begin{align*}
\frac{\partial u_1}{\partial t_1}= &\theta(u_2)- u_2,\\
\frac{\partial u_2}{\partial t_1}= &(\partial_qu_2)\theta^{-1}+\theta(u_3)+u_1u_2-u_2\theta^{-1}(u_1)- u_3,\\
\frac{\partial u_3}{\partial t_1}= &(\partial_qu_3)\theta^{-1}+\theta(u_4)+u_1u_3+u_2(\theta^{-2}(\partial_qu_1))\theta^{-1}
-u_3\theta^{-2}(u_1)- u_4,\\
\frac{\partial u_4}{\partial t_1}= &(\partial_qu_4)\theta^{-1}+\theta(u_5)+u_1u_4-u_2(\theta^{-3}(\partial_q^2u_1))\theta^{-2}
-u_4\theta^{-3}(u_1)- u_5 \\
&+(2)_qu_3(\theta^{-3}(\partial_qu_1))\theta^{-1}.
\end{align*}

\vspace{8pt}
The Lax operator $L$ in
eq.(\ref{qkplaxoperator}) can be generated by  a pseudo-difference operator
$S=1+ \sum_{k=1}^{\infty}s_k
\partial_q^{-k}$ in the following way
\begin{equation}\label{qkpdressingop}
L=S  \partial_q  S^{-1}.
\end{equation}
Here  $S$ is called  dressing operator or wave operator of the  symmetric $q$-KP hierarchy.

 \vspace{8pt}

 \noindent \textbf{Proposition 3.}
Dressing operator $S$  of the  symmetric $q$-KP hierarchy  satisfies the Sato equation
\begin{equation}\label{qkpSatoequation}
\dfrac{\partial S}{\partial t_j}=-(L^j)_-S, \quad j=1,2, 3, \cdots.
\end{equation}

\noindent \textbf{Proof.}  From the Lax equation  $\dfrac{\partial L}{\partial t_n}=[B_n, L]$, which is followed by
\begin{align*}
\dfrac{\partial L}{\partial t_j}&=[B_j, L]=(L^j)_+L-L(L^j)_+\\
&=(L^j-(L^j)_-)L-L(L^j-(L^j)_-) \\
&=-(L^j)_-L+L(L^j)_-.
\end{align*}
On the other hand,
\begin{align*}
\dfrac{\partial L}{\partial t_j}&=\dfrac{\partial }{\partial t_j}(S  \partial_q S^{-1})\\
&=\dfrac{\partial S}{\partial t_j} \partial_q  S^{-1}+S  \partial_q  \dfrac{\partial S^{-1}}{\partial t_j} \\
&=\dfrac{\partial S}{\partial t_j} S^{-1}S \partial_q S^{-1}+S \partial_q (-S^{-1}  \dfrac{\partial S}{\partial t_j} S^{-1}) \\
&=\dfrac{\partial S}{\partial t_j}  S^{-1} L- L \dfrac{\partial S}{\partial t_j} S^{-1},
\end{align*}
then
\begin{equation*}
\dfrac{\partial L}{\partial t_j}=-(L^j)_-L+L(L^j)_-=\dfrac{\partial S}{\partial t_j}  S^{-1} L- L \dfrac{\partial S}{\partial t_j} S^{-1}.
\end{equation*}
The above equation implies that
\begin{equation*}
\dfrac{\partial S}{\partial t_j}S^{-1}=-(L^j)_-, \quad j=1,2, 3, \cdots,
\end{equation*}
which ends the proof.
 \hfill $\square$

\vspace{8pt}

 \noindent \textbf{Definition 1.}
The $q$-wave function $w_q(x,t;z)$  for  the
 symmetric $q$-KP hierarchy  eq.(\ref{qkplaxequ}) with the  wave operator  $S$ in  eq.(\ref{qkpdressingop})
is given  by

\begin{equation}\label{qkpwavefun}
w_q(x,t;z)=S e_q(xz) \exp({\sum  _{i=1}^{\infty}t_iz^i}),
\end{equation}
where $t=(t_1,t_2,t_3,\cdots)$.

\vspace{8pt}

 \noindent \textbf{Proposition 4.}
The $q$-wave function $w_q(x,t;z)$   of the symmetric $q$-KP hierarchy satisfies the following linear $q$-differential
equations
\begin{align*}
Lw_q &=zw_q, \partial_{m}w_q=(L^m)_+w_q,
\end{align*}
where $\partial_{m}=\frac{\partial}{\partial{t_m}}$.

\noindent \textbf{Proof.} Using the equation $\partial_q e_q(xz)=ze_q(xz)$, then
\begin{align*}
Lw_q &=S\partial_q  S^{-1} S e_q(xz) \exp({\sum  _{i=1}^{\infty}t_iz^i})\\
&=S\partial_q e_q(xz) \exp({\sum  _{i=1}^{\infty}t_iz^i})\\
&=zw_q.
\end{align*}

From the Sato equation  $\partial_{m}S=-(L^m)_-S$,  it follows that
\begin{align*}
\partial_{m}w_q&=\partial_{m}(S e_q(xz) \exp({\sum  _{i=1}^{\infty}t_iz^i})) \\
&=(\partial_{m}S) e_q(xz) \exp({\sum  _{i=1}^{\infty}t_iz^i})+S e_q(xz) \exp({\sum  _{i=1}^{\infty}t_iz^i})z^m \\
&=-(L^m)_-Se_q(xz) \exp({\sum  _{i=1}^{\infty}t_iz^i})+S\partial_q^m e_q(xz) \exp({\sum  _{i=1}^{\infty}t_iz^i}) \\
&=-(L^m)_-w_q+(L^m)_+w_q \\
&=(L^m)_+w_q.
\end{align*}
 \hfill $\square$

\vspace{8pt}

 Furthermore,  we  would like to give the definitions  of the
 symmetric $q$-CKP hierarchy and the symmetric $q$-BKP hierarchy respectively to answer the
 previous question mentioned in the introduction.

\vspace{8pt}

 \noindent \textbf{Definition 2.} Let the
operator $L$ in  eq.(\ref{qkplaxoperator}) is the Lax operator for the symmetric $q$-KP hierarchy associated with
 eq.(\ref{qkplaxequ}),  if $L$ satisfies the reduction condition $L^*=-L$, then  we call it the symmetric $q$-CKP hierarchy.

\vspace{8pt}

 \noindent \textbf{Definition 3.} Let the
operator $L$ in  eq.(\ref{qkplaxoperator}) is the Lax operator for the symmetric $q$-KP hierarchy associated with
 eq.(\ref{qkplaxequ}),  if $L$ satisfies the reduction condition $L^*=-\theta^{-\frac{1}{2}}\partial_q L
\partial_q^{-1}\theta^{\frac{1}{2}}$, then it is the symmetric $q$-BKP hierarchy.

\vspace{8pt}

\section{Additional symmetry of  the symmetric  $q$-KP hierarchy}

The another main goal of this note is to consider the  additional symmetries of the symmetric  $q$-KP hierarchy. First, let us
 define $\Gamma_q$ and Orlov-Shulman's $M$ operator  as
\begin{align*}
\Gamma_q &=\sum_{i=1}^{\infty}\Big(it_i+ic_ix^i\Big)\partial
_q^{i-1}, \\
M &= S \Gamma_q S^{-1},
\end{align*}
where $c_i$ is given by eq.(\ref{ck}). Then the additional flows of  the symmetric
 $q$-KP hierarchy for each pair \{$m,n$\} are
defined by
\begin{equation}\label{additionalsym}
\dfrac{\partial S}{\partial t_{m,n}^*}=-(M^mL^n)_-S.
\end{equation}

\vspace{8pt}

 \noindent \textbf{Proposition 5.} The additional flows act on $L$ and $M$ of the symmetric
 $q$-KP hierarchy as
\begin{align}
\dfrac{\partial L}{\partial t_{m,n}^*}&=-[(M^mL^n)_-,L], \\
\dfrac{\partial M}{\partial t_{m,n}^*}&=-[(M^mL^n)_-,M].
\end{align}

\noindent \textbf{Proof.} By performing the derivative $\dfrac{\partial}{\partial t_{m,n}^*}$ on $L=S  \partial_q  S^{-1}$ and using the  eq.(\ref{additionalsym}), we observe that
\begin{align*}
\dfrac{\partial L}{\partial t_{m,n}^*}&= \dfrac{\partial S}{\partial t_{m,n}^*}\partial_q  S^{-1}  + S \partial_q \dfrac{\partial  S^{-1} }{\partial t_{m,n}^*} \\
&=-(M^mL^n)_-S\partial_q  S^{-1}+S \partial_q (-S^{-1}    \dfrac{\partial S}{\partial t_{m,n}^*}        S^{-1}) \\
&=-(M^mL^n)_-L+S \partial_qS^{-1}(M^mL^n)_-\\
&=-[(M^mL^n)_-,L].
\end{align*}
For the action on $M=S \Gamma_q S^{-1}$, there exists similar derivation as $\dfrac{\partial L}{\partial t_{m,n}^*}$, and then
\begin{align*}
\dfrac{\partial M}{\partial t_{m,n}^*}&= \dfrac{\partial S}{\partial t_{m,n}^*}\Gamma_q  S^{-1}  + S \Gamma_q \dfrac{\partial  S^{-1} }{\partial t_{m,n}^*} \\
&=-(M^mL^n)_-S\Gamma_q  S^{-1}+S \Gamma_q (-S^{-1}    \dfrac{\partial S}{\partial t_{m,n}^*}        S^{-1}) \\
&=-(M^mL^n)_-M+S \Gamma_qS^{-1}(M^mL^n)_-\\
&=-[(M^mL^n)_-,M].
\end{align*}
In the above calculation, the fact that $\Gamma_q$ does not depend on the additional flows variables $t_{m,n}^*$ has
been used. \hfill $\square$

 \vspace{8pt}

 \noindent \textbf{ Corollary 1.}
 \begin{align}
\dfrac{\partial L^k}{\partial t_{m,n}^*}&=-[(M^mL^n)_-,L^k], \\
\dfrac{\partial M^k}{\partial t_{m,n}^*}&=-[(M^mL^n)_-,M^k],\\
\dfrac{\partial M^kL^l}{\partial t_{m,n}^*}&=-[(M^mL^n)_-,M^kL^l],\\
\dfrac{\partial M^kL^l}{\partial t_{n}}&=[B_n,M^kL^l].
\end{align}

 \noindent \textbf{Proof.} We present  only the proof of the first equation here. The others can be
proved in a similar way.
  \begin{align*}
\dfrac{\partial L^k}{\partial t_{m,n}^*}&= \dfrac{\partial L}{\partial t_{m,n}^*}L^{k-1}+L\dfrac{\partial L}{\partial t_{m,n}^*}L^{k-2}
+\cdots +L^{k-2}\dfrac{\partial L}{\partial t_{m,n}^*}L+L^{k-1}\dfrac{\partial L}{\partial t_{m,n}^*} \\
&=\sum_{l=1}^k L^{l-1}\dfrac{\partial L}{\partial t_{m,n}^*}L^{k-l} \\
&=\sum_{l=1}^k L^{l-1} (- [(M^mL^n)_-,L])  L^{k-l} \\
&=-[(M^mL^n)_-,L^k],
\end{align*}
where we have used the formula $\dfrac{\partial L}{\partial t_{m,n}^*}=-[(M^mL^n)_-,L]$ in the Proposition 5.  \hfill $\square$

 \vspace{8pt}

  \noindent \textbf{Proposition 6.}
The additional flows ${\partial_{mn}^*}= \dfrac{\partial }{\partial
t_{m,n}^*}$  commute with the hierarchy $\partial_k=\dfrac{\partial
}{\partial t_k}$, i.e. $$[\partial_{mn}^*,\partial_k]=0,$$ thus we call them additional  symmetries of the symmetric
 $q$-KP hierarchy.

 \noindent \textbf{Proof.}
According to the definition and the Corollary 1, it equals to
  \begin{align*}
  [\partial_{mn}^*,\partial_k]S &=\partial_{mn}^*(\partial_kS)-\partial_k(\partial_{mn}^*S) \\
  &=\partial_{mn}^*( -(L^k)_-S)-\partial_k(-(M^mL^n)_-S) \\
  &=-(\partial_{mn}^*L^k)_-S-(L^k)_-(\partial_{mn}^* S)+(\partial_kM^mL^n)_-S+(M^mL^n)_-(\partial_kS)\\
  &=[(M^mL^n)_-,L^k]_-S+(L^k)_- (M^mL^n)_-S +[(L^k)_+,M^mL^n]_-S-(M^mL^n)_-(L^k)_-S\\
  &=[(M^mL^n)_-,L^k]_-S-[(M^mL^n)_-, (L^k)_+]S +[(L^k)_-,(M^mL^n)_-]S\\
  &=[(M^mL^n)_-,(L^k)_-]_-S+[(L^k)_-,(M^mL^n)_-]S
 \\& =0.
  \end{align*}
 $[(L^k)_+,(M^mL^n)]_-=[(L^k)_+,(M^mL^n)_-]_-$ and $[(M^mL^n)_-,(L^k)_-]_-=[(M^mL^n)_-,(L^k)_-]$  have been used in the  above derivation.
\hfill $\square$

\section{Conclusions and discussions}
To summarize, we have derived  the antisymmetric property of $\partial_q$ in Proposition 1 and  a
crucial expression of  $e_q(x)$ by usual exponential in Proposition 2. The analytic property of
symmetric $e_q(x)$ in Proposition 2 is used to define  the wave function of the symmetric $q$-KP
hierarchy. After introducing  the dressing operator and the $q$-wave function of the symmetric $q$-KP
hierarchy in Section 3, we also give the definitions of symmetric $q$-CKP hierarchy and symmetric
$q$-BKP hierarchy. The additional symmetries of the symmetric $q$-KP hierarchy are obtained in
Section 4. The above results of this paper show obviously that the symmetric $q$-KP hierarchy is
different with the $q$-KP hierarchy\cite{ilive3,mas,iliev1,iliev2,tu,he,hetianqkp,hetianqkp2,
linrunliang1,linrunliang2} based on the $D_q(f(x))$.

 \vspace{8pt}
In comparison with the known interesting results of the KP hierarchy~\cite{dkjm,dkjm2,dickey2} and
the $q$-KP hierarchy based on the $D_q(f(x))$ \cite{ilive3,mas,iliev1,iliev2,tu,he,hetianqkp,hetianqkp2,
linrunliang1,linrunliang2}, the symmetric $q$-KP hierarchy defined in this paper deserves further
study from several aspects including the tau function and its Hirota bilinear identity,
the Hamiltonian structure, the gauge transformation, the symmetry analysis and the interaction of
$q$-solitons. Furthermore, it is highly nontrivial to consider above topics of the
symmetric $q$-CKP(or $q$-BKP) hierarchy because of the reduction condition $L^*=-L$(or
$L^*=-\partial_q L \partial_q^{-1}$ ) and the complexity of the $\partial_q$.
\vspace{12pt}

{\bf Acknowledgments} {\small  This work was supported by Erasmus Mundus Action 2 EXPERTS,  SMSTC grant no. 12XD1405000, Fundamental Research Funds for the Central Universities, and NSF grant no. 11271210, 11201451, 10825101 of China.
 }



\end{document}